\documentclass[aps,pre,twocolumn,showpacs,superscriptaddress, floatfix]{revtex4-1}  
\usepackage{amsmath}
\usepackage{amsfonts}
\usepackage{amssymb}
\usepackage{graphicx}

\graphicspath{{./}{figures/}}

\begin{document}


\title{Swarming of Self-Propelled Camphor Boats}


\author{Eric Heisler}
\affiliation{Department of Mathematical and Life Sciences, Hiroshima University, 1-3-1 Kagamiyama, Higashi-Hiroshima 739-8526, Japan}
\author{Nobuhiko J. Suematsu}
\affiliation{Graduate School of Advanced Mathematical Sciences, Meiji University, 1-1-1 Higashimita, Tamaku, Kawasaki 214-8571, Japan}
\affiliation{Meiji Institute for Advanced Study of Mathematical Sciences (MIMS), 1-1-1 Higashimita, Tamaku, Kawasaki 214-8571, Japan}
\author{Akinori Awazu}
\affiliation{Department of Mathematical and Life Sciences, Hiroshima University, 1-3-1 Kagamiyama, Higashi-Hiroshima 739-8526, Japan}
\author{ Hiraku Nishimori}
\affiliation{Department of Mathematical and Life Sciences, Hiroshima University, 1-3-1 Kagamiyama, Higashi-Hiroshima 739-8526, Japan}

\date{\today}

\begin{abstract}
When an ensemble of self-propelled camphor boats move in a one-dimensional channel, they exhibit a variety of collective behaviors. Under certain conditions, the boats tend to cluster together and move in a relatively tight formation. This type of behavior, referred to as clustering or swarming here, is one of three types recently observed in experiment. Similar clustering behavior is also reproduced in simulations based on a simple theoretical model. Here we examine this model to determine the clustering mechanism and the conditions under which clustering occurs. We also propose a method of quantifying the behavior that may be used in future experimental work.
\end{abstract}


\maketitle

\section{Introduction}
Ensembles of many motile, interacting bodies may exhibit a variety of organized phenomena. One such behavior is the grouping of several individuals into a spatially compact formation, which is maintained as the bodies move. We refer to this type of collective motion as clustering. It has been observed in the motion of ants~\cite{Nishinari2006, John2009} and numerous other biological systems~\cite{Okubo1986, Vicsek2000, Vicsek2010} as well as some artificial systems~\cite{Soh2011}. In many cases, the individual members of the group do not interact directly, but rather change their surroundings in ways that influence the behavior of other members. Camphor boats belong to this category because, aside from collisions, they interact only through the background camphor field. Another example of particular relevance to this work is the pheromone trail created by an ant, which signals other ants to follow. The experimental observations of John $et$ $al$.~\cite{John2009} included moving clusters of ants, which they labeled platoons. These platoons share many characteristics with the clustering of camphor boats as described further below.

Camphor boats provide a simple way to experimentally examine collective motion of self-propelled elements. Recent camphor boat experiments have displayed various types of collective behavior such as two-boat interactions~\cite{Kohira2001, Nakata2005} and traffic jams~\cite{Suematsu2010}. Two-dimensional systems of camphor particles have also displayed pattern formation~\cite{Soh2008, Soh2011}, but these systems are fundamentally different in that they contain free camphor particles with no directional orientation, whereas camphor boats are light, plastic disks with attached camphor particles that are driven in a specified direction by surface tension gradients.

\begin{figure}
	\includegraphics[width=8.6cm]{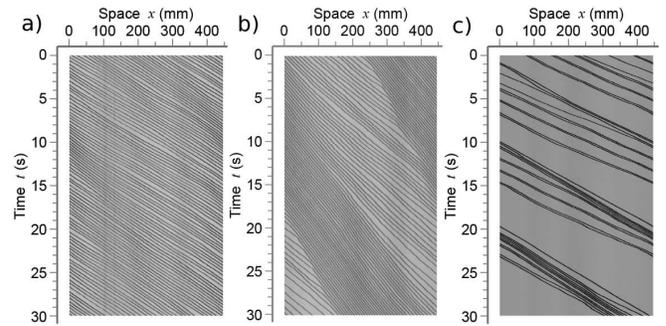}
	\caption{Space-time diagrams from experimental data showing the three modes of collective behavior: a)homogeneous(35 boats, $25^{\circ}$C) b)jammed(40 boats, $25^{\circ}$C) c)clustered(10 boats, $50^{\circ}$C). Black lines indicate the trajectory of boats.}
	\label{fig:three_modes}
\end{figure}

In previous work~\cite{Suematsu2010} we presented three types of collective behavior. Primarily, the boats displayed either roughly homogeneous, constant velocity flow or congested flow(Fig.\ref{fig:three_modes}a and \ref{fig:three_modes}b) depending on the number of boats in the system. The latter occurs with a larger number of boats, and is characterized by periodic acceleration and deceleration. In addition, we observed a mode in which the boats gathered together into high density groups that were maintained over time as shown in Fig.\ref{fig:three_modes}c. This clustering behavior was briefly described therein, but the mechanism and necessary conditions for clustering remained under investigation. The theoretical analysis presented here reveals the mechanism and necessary conditions, and provides a method of quantifying clustering. In this paper, we will describe this theoretical mechanism for clustering and present the results of numerical simulations.

We will define clustering for this study as the behavior in which two or more moving boats gather into a relatively close formation and remain in that formation as they continue moving. This is distinct from the congested behavior previously studied~\cite{Suematsu2010} in that the boats do not experience periodic acceleration, and the formation persists for a long time. Also, the phenomenon can occur with a small total density, whereas congestion typically occurs under higher densities.

\section{Model}
There have been several similar models proposed to describe the motion of a camphor boat ~\cite{Nagayama2004, Suematsu2010, Kohira2001, Hayashima2001}. These models share the same basic concepts, but differ slightly in the mathematical details. The model presented here borrows ideas from these. The camphor boats are thin, circular, plastic disks with a smaller camphor pellet attached to the underside so that half of the pellet overlaps the edge of the disk and half extends beyond it. We will name this overlapping region the back of the boat, and the opposite side is the front. All boats are oriented in the same direction. If the boats are confined to a long, narrow channel in which they cannot pass each other or turn around, the motion is essentially one-dimensional. The 1-d equations of motion of a boat are given by eq.\ref{eq:model_v1}.
\begin{equation}
m\frac{\partial ^{2}x_{i}}{\partial t^{2}} = -\mu \frac{\partial x_{i}}{\partial t} + L\left[ \gamma \left(c(x_{i}+L)\right) - \gamma \left(c(x_{i})\right)\right]
\label{eq:model_v1}
\end{equation}
where $m$ is the mass of the boat, $L$ is the diameter, and $\mu$ is the viscosity constant of the water. The position, $x_{i}$, is defined as the back edge of the boat, which is also the center of the camphor pellet. The second term on the right represents the difference in surface tension between the front and back of the boat as a function of the camphor concentration, $c(x_{i}+L)$ and $c(x_{i})$ respectively. $\gamma(c)$ is typically approximated by a sigmoidal function. We will use the relation in eq.\ref{eq:model_gamma}.
\begin{equation}
\gamma(c) =\frac{\gamma_{water} - \gamma_{camphor}}{\left(\beta c\right)^{2} +1} + \gamma_{camphor}
\label{eq:model_gamma}
\end{equation}
where $\gamma_{water}=72g/s^{2}$ and $\gamma_{camphor}=50g/s^{2}$ are the surface tension of pure water and a camphor saturated solution respectively. The point $c=1/\beta$ is the midpoint between $\gamma_{water}$ and $\gamma_{camphor}$.

The concentration of camphor on the surface of the water is constantly changing due to several processes. For a system with $N$ boats, it can be approximated by the following reaction-diffusion equation \ref{eq:model_c}.
\begin{eqnarray}
\frac{\partial c}{\partial t} = D\frac{\partial^{2}c}{\partial x^{2}} - kc + \alpha \sum_{i=1}^{N} F(x-x_{i}) \label{eq:model_c} \\
F(x) = 1 \ \rm{: for} \ \left| x \right| \leq r_{0}, \ 0 \ \rm{: otherwise} \nonumber
\end{eqnarray}
Here, $D$ is the diffusion constant, $k$ is a constant combining the effects of evaporation and dissolution, and $\alpha F(x-x_{i})$ represents the addition of camphor by each boat's pellet which is centered at the point $x_{i}$ and has radius $r_{0}$. To non-dimensionalize the problem, first define the following dimensionless quantities.
\begin{equation}
t'= t \frac{D}{L^{2}} \ , \ x'=\frac{x}{L} \ , \ c'= c \beta
\label{eq:nondimquant}
\end{equation}
Then the dimensionless parameters of the system become
\begin{eqnarray}
\mu'=\frac{\mu L^{2}}{m D} \ , \ k' = \frac{k L^{2}}{D} & , \ & \Gamma=\frac{L^{3}(\gamma_{w} - \gamma_{c})}{m D^{2}} \\ \nonumber
r_{0}' = \frac{r_{0}}{L} \ , \ R' = \frac{R}{L} & , \ & \alpha'=\frac{\alpha \beta L^{2}}{D}
\label{eq:nondimparam}
\end{eqnarray}
Dropping the $'$ marks, the non-dimensional equations are
\begin{eqnarray}
\frac{\partial ^{2}x_{i}}{\partial t^{2}} & = & -\mu \frac{\partial x_{i}}{\partial t}  + \Gamma \left[ \frac{1}{c(x_{i}+1)^{2} + 1} - \frac{1}{c(x_{i})^{2} + 1} \right]  \\
\frac{\partial c}{\partial t} & = & \frac{\partial^{2}c}{\partial x^{2}} - kc + \alpha \sum_{i=1}^{N} F(x-x_{i}) \\
\ & \ & F(x) = 1 \ \rm{: for} \ \left| x \right| < r_{0}, \ 0 \ \rm{: otherwise} \nonumber 
\end{eqnarray}

\section{Analysis}
\begin{figure}
	\includegraphics[width=8.6cm]{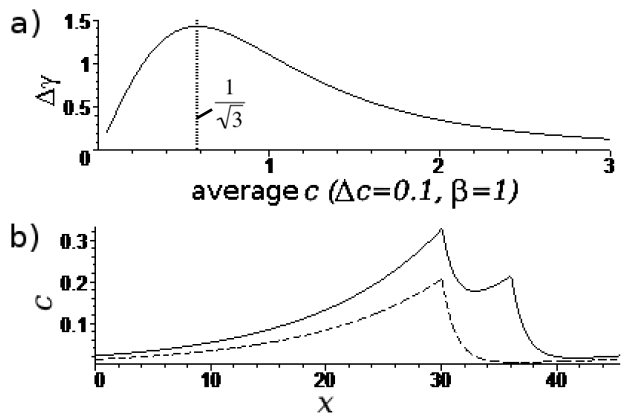}
	\caption{a) The difference in surface tension increases with increasing average $c$ up to $c_{t}$. b) Steady state $c(x)$ for one boat(dashed) and two boats(solid) moving at constant velocity to the right. The peaks approximate the rear edges of the boats, which have length 0.6}
	\label{fig:twoboats}
\end{figure}

The mechanism responsible for clustering can be found in the surface tension function, $\gamma(c)$. For concentrations less than a critical value, $c_{t}= 1/(\sqrt{3}\beta)$, the slope of $\gamma(c)$ becomes steeper with increasing $c$. This means that the difference in $\gamma$ for a fixed difference in $c$ will increase with increasing average $c$ up to $c_{t}$. This is shown in Fig.\ref{fig:twoboats}a. To relate this to a group of boats, if the average camphor level around the leading boat is lower than around trailing boats, the leading boat may feel a weaker driving force. The trailing boats would then move at a higher velocity and approach the leading boat.

We can easily apply this idea to the case of two boats. First, assume that the boats are moving with constant velocity, $v$, and that the camphor concentration in the moving reference frame of the boats has reached a steady state. The time independent value of $c$, in the frame of the boats is given by eq.\ref{eq:oneboat_diff}.
\begin{equation}
\frac{\partial^{2}c}{\partial x^{2}}  -v\frac{\partial c}{\partial x} - kc = - \alpha \sum_{i=1}^{2} F(x-x_{i}) \label{eq:oneboat_diff} 
\end{equation}

The solutions to this equation for one boat and two boats are shown in fig.\ref{fig:twoboats}b. Note the higher concentration around the trailing peak. This feature remains even if the boats are placed close enough to be in contact. By the mechanism described above, the two boats will approach each other and travel in a close, stable configuration. If additional boats are added to the system, they will be gathered into the cluster through the same mechanism. This leads to the question of whether the cluster remains stable as the number of boats increases. 

Using the constant velocity assumption, it is possible to construct a stable cluster with many boats. We can then set the location of the rearmost boat and calculate the net force on it as a function of distance from the next forward boat. This allows us to determine the stability of the cluster.

\begin{figure}
	\includegraphics[width=8.6cm]{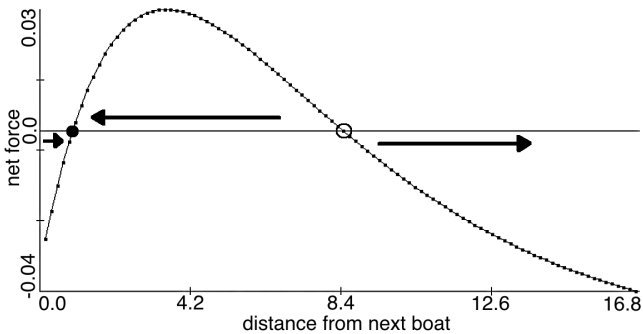}
	\caption{Net force on the last boat vs. its distance from the next forward boat. A positive force will decrease the distance.}
	\label{fig:netforce}
\end{figure}

Under certain conditions, the resulting net force shows a stable equilibrium point at a relatively close distance from the next boat (see fig.\ref{fig:netforce}). The region of attraction, shown as an interval with positive net force, is long compared to the boat length. This makes the equilibrium stable against small perturbations and ensures that boats could be drawn in from an appreciable distance. If the net force remains positive at zero spacing, the boats will remain in contact as they travel. It is important to remember that this mechanism only applies while the average camphor concentration around each boat is less than $c_{t}$. If $c$ is higher than this value, the opposite effect will slow down the trailing boat and effectively spread the boats apart. The periodicity of the system also places a limit on the number of boats that can form a stable cluster as explained below.

Comparing these results with the observations of ants reveals many similarities. As described by John $et$ $al$.~\cite{John2009}, a line of ants may form platoons. These stable, high density groups of ants move at a constant speed. Individual ants following behind the platoon sense its pheromone trail and may move at a higher velocity. This allows the individual ants to catch up with and join the platoon. Clusters of camphor boats are also stable, high density groups that travel at constant speed. The trail of camphor left by the cluster causes following individuals to travel at higher speed and catch up with the cluster. Although camphor boats are not sentient and are driven by much simpler principles, these behaviors are very similar.
\begin{figure}
	\includegraphics[width=8.6cm]{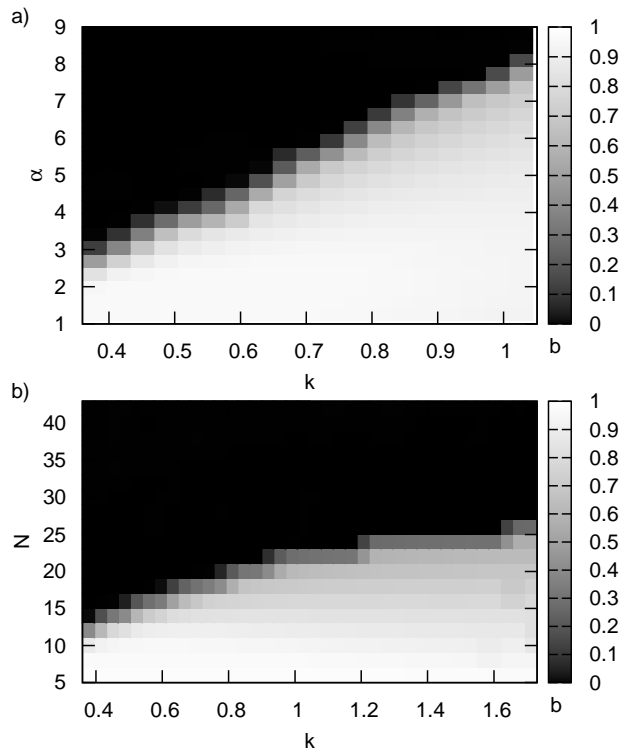}
	\caption{a) $b$ changes abruptly at particular values of $k$ and $\alpha$. b) There is a maximum value of $N$, dependent on various parameters, above which clusters do not form. The other dimensionless parameters used here are $\mu=0.24$, $\Gamma=528$, $R=75.8$, $r_{0}=0.25$.}
	\label{fig:bfigs}
\end{figure}

\section{Numerical results}
It is important to provide a quantitative way to discern clustering from other kinds of behavior. Considering the structure and stability of clusters, we propose the following quantity as an order parameter.
\begin{equation}
b = \displaystyle\max_{i=1..N} \frac{\langle d_{fi} - d_{bi} \rangle }{R-NL}
\label{eq:b_def}
\end{equation}
For boat number $i$, $d_{fi}$ and $d_{bi}$ are the distances from the next boat in front and in back respectively, and the normalizing factor, $R-NL$, is the total empty space in the route. The brackets represent a time average. For a system containing one cluster with all boats in contact, $b=1$. For ideal homogeneous flow, $b=0$. For jammed flow, $b$ will approach zero when averaging over long time intervals. In practice, particularly in experiment, $b$ may deviate slightly from these values, but should generally be much larger for the clustered mode. 

By testing the dependence of $b$ on the model parameters, we found that the most important dependence was on the parameters related to the camphor level. These include the evaporation constant, $k$, and the supply parameters, $\alpha$ and $r_{0}$. The relation to $k$ and $\alpha$ can be seen in Fig.\ref{fig:bfigs}a. Since the effect of $r_{0}$ is very similar to that of $\alpha$, and its values are very limited in practice, it was not examined separately.

There was also a dependence on the number of boats. One reason is that adding more boats will increase the total camphor being added to the water. Also, if the size of the cluster becomes comparable to the total route length, the boats at the back of the cluster will influence the boats at the front. In simulation this effect disrupted the stability and destroyed the cluster. Considering these two effects, there appears to be a maximum number of boats above which clusters do not form as shown in fig.\ref{fig:bfigs}. This maximum number depends on several factors, but our tests suggest that it is typically less than the critical number required for the jammed mode to appear. Thus, for the parameter space tested, clustering and jamming did not coexist. However, it is possible that there exists some set of parameters for which clustering and jamming can both occur.

\begin{figure}
	\includegraphics[width=8.6cm]{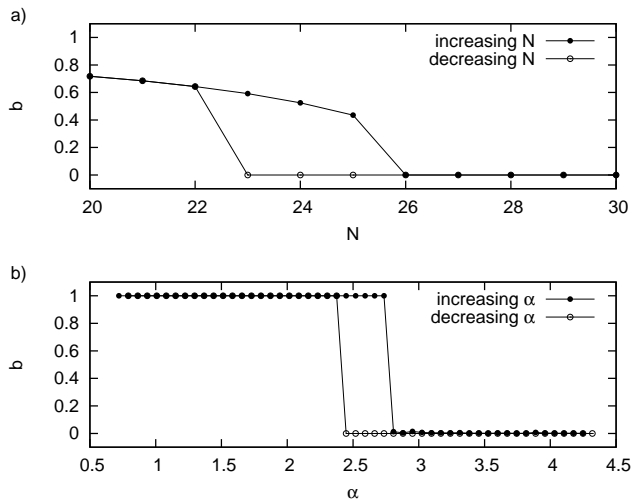}
	\caption{Hysteresis is seen by gradually increasing and decreasing $N$ ($k=0.73$) and $\alpha$ ($k=0.36$). For $k=0.73$, the hysteresis with respect to $\alpha$ vanishes. Other parameters as in Fig.\ref{fig:bfigs}.}
	\label{fig:hyst}
\end{figure}

Considering the very abrupt changes in $b$, it is worthwhile to check for hysteresis. The results shown in fig.\ref{fig:hyst} were obtained by gradually changing either the number of boats $N$ or the supply rate $\alpha$. The system was given time to approach a steady state between each measurement. The figure shows clear hysteresis with respect to the number of boats and the supply rate for this set of parameters, but if $k$ is increased, the hysteresis with respect to $\alpha$ vanishes. This may suggest a transition from a subcritical to a supercritical bifurcation.

\section{Conclusion}
To better understand clustering behavior seen in experiment, we examined a mathematical model of an ensemble of camphor boats in a periodic, one-dimensional system. We found that this behavior was caused by the relationship between surface tension and camphor concentration on the surface of water. For low camphor concentrations, boats may experience a stronger driving force by traveling close behind other boats. We then defined an order parameter to quantify the clustering behavior and applied it to the results of numerical simulations to examine the dependence on system parameters. Our next goal is to investigate this phenomenon experimentally to test the validity of the theoretical analysis presented here. Clustering behavior can be reproduced in experiment, but the dependence on experimental conditions has not yet been determined. 

In analyzing this model of collective motion, we hope to have added some insight into clustering or swarming behavior. The striking similarity to the behavior of ants seems to support the idea that collective motion can be categorized into a set of universal classes. Although the mechanism presented is specific to the motion of camphor boats, it represents yet another way in which nature forms organized structures from relatively simple pieces.

\begin{acknowledgments}
The authors thank Prof. S. Nakata (Hiroshima Univ.) for his helpful suggestions.  This work was supported by Grants-in-Aid for Scientific Research C (No. 22540391) to H.N. and for Young Scientists B (No. 23740299) to N.J.S. from the JPSJ of Japan, and the Global COE Program Formation and Development of Mathematical Sciences Based on Modeling and Analysis from the MEXT of Japan.
\end{acknowledgments}

\bibliography{camphorRefs}

\begin{thebibliography}{12}%
\makeatletter
\providecommand \@ifxundefined [1]{%
 \@ifx{#1\undefined}
}%
\providecommand \@ifnum [1]{%
 \ifnum #1\expandafter \@firstoftwo
 \else \expandafter \@secondoftwo
 \fi
}%
\providecommand \@ifx [1]{%
 \ifx #1\expandafter \@firstoftwo
 \else \expandafter \@secondoftwo
 \fi
}%
\providecommand \natexlab [1]{#1}%
\providecommand \enquote  [1]{``#1''}%
\providecommand \bibnamefont  [1]{#1}%
\providecommand \bibfnamefont [1]{#1}%
\providecommand \citenamefont [1]{#1}%
\providecommand \href@noop [0]{\@secondoftwo}%
\providecommand \href [0]{\begingroup \@sanitize@url \@href}%
\providecommand \@href[1]{\@@startlink{#1}\@@href}%
\providecommand \@@href[1]{\endgroup#1\@@endlink}%
\providecommand \@sanitize@url [0]{\catcode `\\12\catcode `\$12\catcode
  `\&12\catcode `\#12\catcode `\^12\catcode `\_12\catcode `\%12\relax}%
\providecommand \@@startlink[1]{}%
\providecommand \@@endlink[0]{}%
\providecommand \url  [0]{\begingroup\@sanitize@url \@url }%
\providecommand \@url [1]{\endgroup\@href {#1}{\urlprefix }}%
\providecommand \urlprefix  [0]{URL }%
\providecommand \Eprint [0]{\href }%
\@ifxundefined \urlstyle {%
  \providecommand \doi  [0]{\begingroup \@sanitize@url \@doi}%
  \providecommand \@doi [1]{\endgroup \@@startlink {\doibase
  #1}doi:\discretionary {}{}{}#1\@@endlink }%
}{%
  \providecommand \doi  [0]{doi:\discretionary{}{}{}\begingroup
  \urlstyle{rm}\Url }%
}%
\providecommand \doibase [0]{http://dx.doi.org/}%
\providecommand \Doi [0]{\begingroup \@sanitize@url \@Doi }%
\providecommand \@Doi  [1]{\endgroup\@@startlink{\doibase#1}\@@Doi}%
\providecommand \@@Doi [1]{#1\@@endlink}%
\providecommand \selectlanguage [0]{\@gobble}%
\providecommand \bibinfo  [0]{\@secondoftwo}%
\providecommand \bibfield  [0]{\@secondoftwo}%
\providecommand \translation [1]{[#1]}%
\providecommand \BibitemOpen [0]{}%
\providecommand \bibitemStop [0]{}%
\providecommand \bibitemNoStop [0]{.\EOS\space}%
\providecommand \EOS [0]{\spacefactor3000\relax}%
\providecommand \BibitemShut  [1]{\csname bibitem#1\endcsname}%
\bibitem [{\citenamefont {Nishinari}\ \emph {et~al.}(2006)\citenamefont
  {Nishinari}, \citenamefont {Sugawara}, \citenamefont {Kazama}, \citenamefont
  {Schadschneider},\ and\ \citenamefont {Chowdhury}}]{Nishinari2006}%
  \BibitemOpen
  \bibfield  {author} {\bibinfo {author} {\bibfnamefont {K.}~\bibnamefont
  {Nishinari}}, \bibinfo {author} {\bibfnamefont {K.}~\bibnamefont {Sugawara}},
  \bibinfo {author} {\bibfnamefont {T.}~\bibnamefont {Kazama}}, \bibinfo
  {author} {\bibfnamefont {A.}~\bibnamefont {Schadschneider}}, \ and\ \bibinfo
  {author} {\bibfnamefont {D.}~\bibnamefont {Chowdhury}},\ }\Doi
  {10.1016/j.physa.2006.05.016} {\bibfield  {journal} {\bibinfo  {journal}
  {Physica A: Statistical Mechanics and its Applications},\ }\textbf {\bibinfo
  {volume} {372}},\ \bibinfo {pages} {132 } (\bibinfo {year} {2006})},\ ISSN
  \bibinfo {issn} {0378-4371}\BibitemShut {NoStop}%
\bibitem [{\citenamefont {John}\ \emph {et~al.}(2009)\citenamefont {John},
  \citenamefont {Schadschneider}, \citenamefont {Chowdhury},\ and\
  \citenamefont {Nishinari}}]{John2009}%
  \BibitemOpen
  \bibfield  {author} {\bibinfo {author} {\bibfnamefont {A.}~\bibnamefont
  {John}}, \bibinfo {author} {\bibfnamefont {A.}~\bibnamefont
  {Schadschneider}}, \bibinfo {author} {\bibfnamefont {D.}~\bibnamefont
  {Chowdhury}}, \ and\ \bibinfo {author} {\bibfnamefont {K.}~\bibnamefont
  {Nishinari}},\ }\Doi {10.1103/PhysRevLett.102.108001} {\bibfield  {journal}
  {\bibinfo  {journal} {Phys. Rev. Lett.},\ }\textbf {\bibinfo {volume}
  {102}},\ \bibinfo {pages} {108001} (\bibinfo {year} {2009})}\BibitemShut
  {NoStop}%
\bibitem [{\citenamefont {Okubo}(1986)}]{Okubo1986}%
  \BibitemOpen
  \bibfield  {author} {\bibinfo {author} {\bibfnamefont {A.}~\bibnamefont
  {Okubo}},\ }\Doi {DOI: 10.1016/0065-227X(86)90003-1} {\bibfield  {journal}
  {\bibinfo  {journal} {Advances in Biophysics},\ }\textbf {\bibinfo {volume}
  {22}},\ \bibinfo {pages} {1 } (\bibinfo {year} {1986})},\ ISSN \bibinfo
  {issn} {0065-227X}\BibitemShut {NoStop}%
\bibitem [{\citenamefont {{Czirok}}\ and\ \citenamefont
  {{Vicsek}}(2000)}]{Vicsek2000}%
  \BibitemOpen
  \bibfield  {author} {\bibinfo {author} {\bibfnamefont {A.}~\bibnamefont
  {{Czirok}}}\ and\ \bibinfo {author} {\bibfnamefont {T.}~\bibnamefont
  {{Vicsek}}},\ }\Doi {10.1016/S0378-4371(00)00013-3} {\bibfield  {journal}
  {\bibinfo  {journal} {Physica A Statistical Mechanics and its Applications},\
  }\textbf {\bibinfo {volume} {281}},\ \bibinfo {pages} {17} (\bibinfo {year}
  {2000})}\BibitemShut {NoStop}%
\bibitem [{\citenamefont {{Vicsek}}\ and\ \citenamefont
  {{Zafiris}}(2010)}]{Vicsek2010}%
  \BibitemOpen
  \bibfield  {author} {\bibinfo {author} {\bibfnamefont {T.}~\bibnamefont
  {{Vicsek}}}\ and\ \bibinfo {author} {\bibfnamefont {A.}~\bibnamefont
  {{Zafiris}}},\ }\href@noop {} {\bibfield  {journal} {\bibinfo  {journal}
  {ArXiv e-prints}} (\bibinfo {year} {2010})},\ \Eprint
  {http://arxiv.org/abs/1010.5017} {arXiv:1010.5017 [cond-mat.stat-mech]}
  \BibitemShut {NoStop}%
\bibitem [{\citenamefont {Soh}\ \emph {et~al.}(2011)\citenamefont {Soh},
  \citenamefont {Branicki},\ and\ \citenamefont {Grzybowski}}]{Soh2011}%
  \BibitemOpen
  \bibfield  {author} {\bibinfo {author} {\bibfnamefont {S.}~\bibnamefont
  {Soh}}, \bibinfo {author} {\bibfnamefont {M.}~\bibnamefont {Branicki}}, \
  and\ \bibinfo {author} {\bibfnamefont {B.~A.}\ \bibnamefont {Grzybowski}},\
  }\Doi {10.1021/jz200180z} {\bibfield  {journal} {\bibinfo  {journal} {The
  Journal of Physical Chemistry Letters},\ }\textbf {\bibinfo {volume} {2}},\
  \bibinfo {pages} {770} (\bibinfo {year} {2011})}\BibitemShut {NoStop}%
\bibitem [{\citenamefont {Kohira}\ \emph {et~al.}(2001)\citenamefont {Kohira},
  \citenamefont {Hayashima}, \citenamefont {Nagayama},\ and\ \citenamefont
  {Nakata}}]{Kohira2001}%
  \BibitemOpen
  \bibfield  {author} {\bibinfo {author} {\bibfnamefont {M.~I.}\ \bibnamefont
  {Kohira}}, \bibinfo {author} {\bibfnamefont {Y.}~\bibnamefont {Hayashima}},
  \bibinfo {author} {\bibfnamefont {M.}~\bibnamefont {Nagayama}}, \ and\
  \bibinfo {author} {\bibfnamefont {S.}~\bibnamefont {Nakata}},\ }\Doi
  {10.1021/la010388r} {\bibfield  {journal} {\bibinfo  {journal} {Langmuir},\
  }\textbf {\bibinfo {volume} {17}},\ \bibinfo {pages} {7124} (\bibinfo {year}
  {2001})}\BibitemShut {NoStop}%
\bibitem [{\citenamefont {Nakata}\ \emph {et~al.}(2005)\citenamefont {Nakata},
  \citenamefont {Doi},\ and\ \citenamefont {Kitahata}}]{Nakata2005}%
  \BibitemOpen
  \bibfield  {author} {\bibinfo {author} {\bibfnamefont {S.}~\bibnamefont
  {Nakata}}, \bibinfo {author} {\bibfnamefont {Y.}~\bibnamefont {Doi}}, \ and\
  \bibinfo {author} {\bibfnamefont {H.}~\bibnamefont {Kitahata}},\ }\Doi
  {10.1021/jp0480605} {\bibfield  {journal} {\bibinfo  {journal} {The Journal
  of Physical Chemistry B},\ }\textbf {\bibinfo {volume} {109}},\ \bibinfo
  {pages} {1798} (\bibinfo {year} {2005})}\BibitemShut {NoStop}%
\bibitem [{\citenamefont {Suematsu}\ \emph {et~al.}(2010)\citenamefont
  {Suematsu}, \citenamefont {Nakata}, \citenamefont {Awazu},\ and\
  \citenamefont {Nishimori}}]{Suematsu2010}%
  \BibitemOpen
  \bibfield  {author} {\bibinfo {author} {\bibfnamefont {N.~J.}\ \bibnamefont
  {Suematsu}}, \bibinfo {author} {\bibfnamefont {S.}~\bibnamefont {Nakata}},
  \bibinfo {author} {\bibfnamefont {A.}~\bibnamefont {Awazu}}, \ and\ \bibinfo
  {author} {\bibfnamefont {H.}~\bibnamefont {Nishimori}},\ }\Doi
  {10.1103/PhysRevE.81.056210} {\bibfield  {journal} {\bibinfo  {journal}
  {Phys. Rev. E},\ }\textbf {\bibinfo {volume} {81}},\ \bibinfo {pages}
  {056210} (\bibinfo {year} {2010})}\BibitemShut {NoStop}%
\bibitem [{\citenamefont {Soh}\ \emph {et~al.}(2008)\citenamefont {Soh},
  \citenamefont {Bishop},\ and\ \citenamefont {Grzybowski}}]{Soh2008}%
  \BibitemOpen
  \bibfield  {author} {\bibinfo {author} {\bibfnamefont {S.}~\bibnamefont
  {Soh}}, \bibinfo {author} {\bibfnamefont {K.~J.~M.}\ \bibnamefont {Bishop}},
  \ and\ \bibinfo {author} {\bibfnamefont {B.~A.}\ \bibnamefont {Grzybowski}},\
  }\Doi {10.1021/jp7111457} {\bibfield  {journal} {\bibinfo  {journal} {The
  Journal of Physical Chemistry B},\ }\textbf {\bibinfo {volume} {112}},\
  \bibinfo {pages} {10848} (\bibinfo {year} {2008})}\BibitemShut {NoStop}%
\bibitem [{\citenamefont {Nagayama}\ \emph {et~al.}(2004)\citenamefont
  {Nagayama}, \citenamefont {Nakata}, \citenamefont {Doi},\ and\ \citenamefont
  {Hayashima}}]{Nagayama2004}%
  \BibitemOpen
  \bibfield  {author} {\bibinfo {author} {\bibfnamefont {M.}~\bibnamefont
  {Nagayama}}, \bibinfo {author} {\bibfnamefont {S.}~\bibnamefont {Nakata}},
  \bibinfo {author} {\bibfnamefont {Y.}~\bibnamefont {Doi}}, \ and\ \bibinfo
  {author} {\bibfnamefont {Y.}~\bibnamefont {Hayashima}},\ }\Doi {DOI:
  10.1016/j.physd.2004.02.003} {\bibfield  {journal} {\bibinfo  {journal}
  {Physica D: Nonlinear Phenomena},\ }\textbf {\bibinfo {volume} {194}},\
  \bibinfo {pages} {151 } (\bibinfo {year} {2004})},\ ISSN \bibinfo {issn}
  {0167-2789}\BibitemShut {NoStop}%
\bibitem [{\citenamefont {Hayashima}\ \emph {et~al.}(2001)\citenamefont
  {Hayashima}, \citenamefont {Nagayama},\ and\ \citenamefont
  {Nakata}}]{Hayashima2001}%
  \BibitemOpen
  \bibfield  {author} {\bibinfo {author} {\bibfnamefont {Y.}~\bibnamefont
  {Hayashima}}, \bibinfo {author} {\bibfnamefont {M.}~\bibnamefont {Nagayama}},
  \ and\ \bibinfo {author} {\bibfnamefont {S.}~\bibnamefont {Nakata}},\ }\Doi
  {10.1021/jp004505n} {\bibfield  {journal} {\bibinfo  {journal} {The Journal
  of Physical Chemistry B},\ }\textbf {\bibinfo {volume} {105}},\ \bibinfo
  {pages} {5353} (\bibinfo {year} {2001})}\BibitemShut {NoStop}%
\end{thebibliography}%

\end{document}